**Title: Electron Diffraction of Water in No Man's Land**

**Authors:** Constantin R. Krüger[1]†, Nathan J. Mowry[1]†, Gabriele Bongiovanni[1]†, Marcel Drabbels[1], and Ulrich J. Lorenz[1]*

**Affiliation:**

[1] Ecole Polytechnique Fédérale de Lausanne (EPFL), Laboratory of Molecular Nanodynamics, CH-1015 Lausanne, Switzerland

* Corresponding author. Email: ulrich.lorenz@epfl.ch

† These authors contributed equally

**Abstract:**

A generally accepted understanding of the anomalous properties of water will only emerge if it becomes possible to systematically characterize water in the deeply supercooled regime, from where the anomalies appear to emanate. This has largely remained elusive because water crystallizes rapidly between 160 K and 232 K. Here, we present an experimental approach to rapidly prepare deeply supercooled water at a well-defined temperature and probe it with electron diffraction before crystallization occurs. We show that as water is cooled from room temperature to cryogenic temperature, its structure evolves smoothly, approaching that of amorphous ice just below 200 K. Our experiments narrow down the range of possible explanations for the origin of the water anomalies and open up new avenues for studying supercooled water.

**One-Sentence Summary:**
Time-resolved electron diffraction captures the structural evolution of water as it is cooled from room temperature to cryogenic temperature.

**Main Text:**

Water has been called "the most anomalous liquid" (*1*), with over 70 anomalous properties that have been identified to date (*2*). Several competing theories have been put forward to explain the origin of these anomalies (*3*, *4*). The liquid-liquid critical point scenario posits that at low temperatures, supercooled water exists in a high and a low density phase, with the phase coexistence line terminating in a critical point (*5*, *6*). In this model, the anomalies manifest as water approaches the Widom line that emanates from this critical point (*7*). The critical-point-free scenario similarly proposes the existence of a liquid-liquid phase transition in the supercooled regime, but with the critical point at negative pressure (*8*, *9*). Other theories explain the anomalies without requiring the existence of a singularity (*10*). Most vexingly, the experimental verification of these theories has largely remained elusive because of the fast crystallization of water in the temperature range of 160–232 K, frequently nicknamed "no man's land" (*3*). X-ray diffraction of evaporatively cooled microdroplets has revealed a smooth evolution of the structure factor down to 227 K (*11*), but has been unable to access lower temperatures due to rapid crystallization. Infrared spectra of transiently heated amorphous ices are consistent with a two state-mixture of a high and a low temperature motif in no man's land (*12*). However, the approach does not probe the liquid itself, but rather an amorphous ice that has sampled a range of temperatures. A definitive explanation of the origin of the water anomalies can only emerge if water can be systematically characterized throughout no man's land. This requires preparing the supercooled liquid at a well-defined temperature and probing it directly before crystallization occurs. Here, we present an experimental approach that overcomes these challenges and allows us to capture the structural evolution of water as it is cooled from room temperature to cryogenic temperature.

Experiments are performed with a time-resolved electron microscope developed in our laboratory (Materials and Methods, sections A-D) (*13*, *14*). The sample geometry and experimental concept are illustrated in Fig. 1. A 600 mesh gold grid (Fig. 1A) supports a holey gold film (2 μm holes) that is covered with a sheet of few-layer graphene (Fig. 1B). The sample is cooled to 101 K, and a 176 nm layer of amorphous solid water is deposited *in situ*. In order to prepare water in no man's land, we locally heat the sample with a shaped microsecond laser pulse, with the laser beam centered onto one of the squares of the specimen grid (532 nm wavelength). We then use an intense, high-brightness electron pulse to capture a diffraction pattern of the supercooled liquid (Fig. 1C).

Figure 2A illustrates the typical shape of the microsecond laser pulse (green) that we use to prepare water in no man's land, with the simulated temperature evolution of the sample shown in black (Materials and Methods, section E). We first heat the sample to room temperature with a 30 μs laser pulse, before reducing the laser power in order to rapidly cool the liquid to a well-defined temperature in no man's land. This sequence maximizes the observation time that is available before crystallization sets in. In contrast, if the sample is simply heated up to reach no man's land, it crystallizes rapidly, since it first has to pass through the temperature regime around 185 K, where the nucleation rate has a maximum (*3*). In the example simulation shown in Fig. 2A, the laser power is reduced by half, which causes the liquid to rapidly cool from 300 K to 236 K. At a delay of 15 μs after reducing the laser power, we then capture a diffraction pattern of the supercooled liquid with a 6 μs electron pulse before crystallization sets in. Note that in order to reduce the cooling time, we initially lower the laser power even further, as shown in Fig. 2A.

As detailed in Supplementary Text, section H, we confirm that the temperature of the sample has stabilized when we probe its structure by characterizing the cooling process with time-resolved electron diffraction. Even in the temperature range where the cooling is expected to be slowest, we measure a 1/e cooling time of only 11 μs. The fast cooling is a consequence of the close proximity of the area under observation to the bars of the specimen grid, which remain at cryogenic temperature throughout the experiment and therefore act as an efficient heat sink (*15–17*).

The specific heat transfer properties of the sample geometry make it straightforward to determine the temperature at which the liquid stabilizes for a given laser power (Materials and Methods, section F). Simulations show that this temperature increases linearly with laser power and only starts to level off above ~260 K, where evaporative cooling becomes important (Fig. 2B). This allows us to determine the sample temperature as follows. By comparing with x-ray data in the mildly supercooled regime (*11, 18*), we obtain a temperature calibration of our experiment at intermediate laser powers. At zero laser power, the sample temperature only slightly exceeds the temperature before the laser pulse. We can therefore use simulations to determine the temperature at zero laser power with good accuracy. For all other laser powers in the linear regime (which includes no man's land), we interpolate linearly. Data recorded at higher laser powers are corrected for the effect of evaporative cooling as described in Materials and Methods, section F.

Figure 3A shows the temperature evolution of the diffraction pattern of water, revealing that the structure of water evolves smoothly as the liquid is cooled from 290 K to 180 K (Materials and Methods, section G). This is also evident in the two-dimensional plot of Fig. 3B, where the positions of the first two diffraction maxima are indicated with black dots. The grey lines represent splines that provide a guide to the eye. As shown in Fig. S9, the positions of the diffraction maxima are largely consistent with x-ray data that are available for temperatures above 227 K (*11, 18*). The position of the second diffraction maximum exhibits a somewhat smaller temperature dependence in our experiment, which is likely due to an overlap with a diffraction feature arising from the graphene support. We therefore do not include the evolution of the second diffraction maximum in the interpretation of our data.

The position of the first diffraction maximum of water exhibits an s-shaped temperature evolution. Upon cooling the room temperature liquid, the maximum shifts to lower momentum transfer, with the shift accelerating below about 240 K. Just below 200 K, the peak position converges to that of hyperquenched glassy water (HGW, horizontal lines in Fig. 3B), a form of amorphous ice that is formed when liquid water is cooled at rates exceeding $10^6$ K/s (*19*) and that we obtain in our experiment when we simply switch off the laser to let the liquid cool at maximum speed.

Our data reveal a continuous evolution of the structure of water as the liquid is cooled, which only slows once it approaches the structure of HGW below 200 K. We therefore infer that at higher temperatures, the liquid fully relaxes on the timescale of our experiment. This is consistent with previous studies, which show that the relaxation time of water exceeds the timescale of our experiment (~5 μs) only at temperatures below 185 K (20). Our data thus confirm that water in no man's land can be equilibrated before crystallization occurs, a point that had previously been debated (6, 11, 21).

The smooth evolution of the diffraction pattern of water appears inconsistent with theories that predict the liquid to undergo a first-order phase transition under our experimental conditions. In this case, one would expect to observe a signature in the temperature dependence of the structure factor (22). Instead, the evolution of water from its room-temperature to its low-temperature structure occurs continuously over a wide temperature interval of about 40 K, between 220 K and 260 K.

In the liquid-liquid critical point scenario, the anomalies of water arise as the liquid approaches the Widom line, which represents the locus of the maximum rate of change of its properties. Studies on transiently heated ice films have placed this point at 210 K (23), while x-ray diffraction experiments on microdroplets have deduced a temperature of 229 K (18). Our analysis has the benefit of including the entire temperature range. Interestingly, it places the temperature of maximum change at 243±2 K, where the evolution of the first diffraction maximum has an inflection point.

By capturing the structural evolution of water throughout no man's land, our experiments help narrow down the range of possible explanations for the origin of the water anomalies and provide a stringent test for the development of accurate water models. Our approach for rapidly preparing water in no man's land and equilibrating it at a well-defined temperature should be quite general. For example, it will be straightforward to combine it with a range of different probes, such as infrared, Raman (24), or x-ray absorption spectroscopy, which are each sensitive to different properties of water. Moreover, our approach opens up new avenues for studying the dynamics of supercooled water, with preliminary experiments showing that we can obtain insights into the crystallization process. Finally, our experiments also bear relevance to cryo-electron microscopy, which appears set to become the preferred method in structural biology (25). In cryo-electron microscopy, vitrified protein samples are prepared through hyperquenching. Our experiments capture the structural evolution of water during vitrification and therefore promise to shed new light onto the question of how well the process is able to preserve the room-temperature structure of proteins (26).

# References

1.  L. G. M. Pettersson, R. H. Henchman, A. Nilsson, Water—The Most Anomalous Liquid. *Chem. Rev.* **116**, 7459–7462 (2016).

2.  K. Amann-Winkel, R. Böhmer, F. Fujara, C. Gainaru, B. Geil, T. Loerting, *Colloquium* : Water's controversial glass transitions. *Rev. Mod. Phys.* **88**, 011002 (2016).

3.  P. Gallo, K. Amann-Winkel, C. A. Angell, M. A. Anisimov, F. Caupin, C. Chakravarty, E. Lascaris, T. Loerting, A. Z. Panagiotopoulos, J. Russo, J. A. Sellberg, H. E. Stanley, H. Tanaka, C. Vega, L. Xu, L. G. M. Pettersson, Water: A Tale of Two Liquids. *Chem. Rev.* **116**, 7463–7500 (2016).

4.  G. Pallares, M. El Mekki Azouzi, M. A. Gonzalez, J. L. Aragones, J. L. F. Abascal, C. Valeriani, F. Caupin, Anomalies in bulk supercooled water at negative pressure. *Proc. Natl. Acad. Sci.* **111**, 7936–7941 (2014).

5.  P. H. Poole, F. Sciortino, U. Essmann, H. E. Stanley, Phase behaviour of metastable water. *Nature.* **360**, 324–328 (1992).

6.  J. C. Palmer, F. Martelli, Y. Liu, R. Car, A. Z. Panagiotopoulos, P. G. Debenedetti, Metastable liquid–liquid transition in a molecular model of water. *Nature.* **510**, 385–388 (2014).

7.  L. Xu, P. Kumar, S. V. Buldyrev, S.-H. Chen, P. H. Poole, F. Sciortino, H. E. Stanley, Relation between the Widom line and the dynamic crossover in systems with a liquid–liquid phase transition. *Proc. Natl. Acad. Sci.* **102**, 16558–16562 (2005).

8.  P. H. Poole, F. Sciortino, T. Grande, H. E. Stanley, C. A. Angell, Effect of Hydrogen Bonds on the Thermodynamic Behavior of Liquid Water. *Phys. Rev. Lett.* **73**, 1632–1635 (1994).

9.  C. A. Angell, Insights into phases of liquid water from study of its unusual glass-forming properties. *Science.* **319**, 582–587 (2008).

10. S. Sastry, P. G. Debenedetti, F. Sciortino, H. E. Stanley, Singularity-free interpretation of the thermodynamics of supercooled water. *Phys. Rev. E.* **53**, 6144–6154 (1996).

11. J. A. Sellberg, C. Huang, T. A. McQueen, N. D. Loh, H. Laksmono, D. Schlesinger, R. G. Sierra, D. Nordlund, C. Y. Hampton, D. Starodub, D. P. DePonte, M. Beye, C. Chen, A. V. Martin, A. Barty, K. T. Wikfeldt, T. M. Weiss, C. Caronna, J. Feldkamp, L. B. Skinner, M. M. Seibert, M. Messerschmidt, G. J. Williams, S. Boutet, L. G. M. Pettersson, M. J. Bogan, A. Nilsson, Ultrafast X-ray probing of water structure below the homogeneous ice nucleation temperature. *Nature.* **510**, 381–384 (2014).

12. L. Kringle, W. A. Thornley, B. D. Kay, G. A. Kimmel, Reversible structural transformations in supercooled liquid water from 135 to 245 K. *Science.* **369**, 1490–1492 (2020).

13. G. Bongiovanni, P. K. Olshin, M. Drabbels, U. J. Lorenz, Intense microsecond electron pulses from a Schottky emitter. *Appl. Phys. Lett.* **116**, 234103 (2020).

14. P. K. Olshin, G. Bongiovanni, M. Drabbels, U. J. Lorenz, Atomic-Resolution Imaging of Fast Nanoscale Dynamics with Bright Microsecond Electron Pulses. *Nano Lett.* **21**, 612–618 (2021).

15. J. M. Voss, O. F. Harder, P. K. Olshin, M. Drabbels, U. J. Lorenz, Rapid melting and revitrification as an approach to microsecond time-resolved cryo-electron microscopy. *Chem. Phys. Lett.* **778**, 138812 (2021).

16. J. M. Voss, O. F. Harder, P. K. Olshin, M. Drabbels, U. J. Lorenz, Microsecond melting and revitrification of cryo samples. *Struct. Dyn.* **8**, 054302 (2021).

17. O. F. Harder, J. M. Voss, P. K. Olshin, M. Drabbels, U. J. Lorenz, Microsecond melting and revitrification of cryo samples: protein structure and beam-induced motion. *Acta Crystallogr. Sect. Struct. Biol.* **78**, 883–889 (2022).

18. K. H. Kim, A. Späh, H. Pathak, F. Perakis, D. Mariedahl, K. Amann-Winkel, J. A. Sellberg, J. H. Lee, S. Kim, J. Park, K. H. Nam, T. Katayama, A. Nilsson, Maxima in the thermodynamic response and correlation functions of deeply supercooled water. *Science.* **358**, 1589–1593 (2017).

19. M. Warkentin, J. P. Sethna, R. E. Thorne, Critical Droplet Theory Explains the Glass Formability of Aqueous Solutions. *Phys. Rev. Lett.* **110**, 015703 (2013).

20. L. Kringle, W. A. Thornley, B. D. Kay, G. A. Kimmel, Structural relaxation and crystallization in supercooled water from 170 to 260 K. *Proc. Natl. Acad. Sci.* **118**, e2022884118 (2021).

21. E. B. Moore, V. Molinero, Structural transformation in supercooled water controls the crystallization rate of ice. *Nature.* **479**, 506–508 (2011).

22. P. Jenniskens, D. F. Blake, Structural Transitions in Amorphous Water Ice and Astrophysical Implications. *Science.* **265**, 753–756 (1994).

23. L. Kringle, W. A. Thornley, B. D. Kay, G. A. Kimmel, Reversible structural transformations in supercooled liquid water from 135 to 245 K. *Science.* **369**, 1490–1492 (2020).

24. F. Perakis, L. De Marco, A. Shalit, F. Tang, Z. R. Kann, T. D. Kühne, R. Torre, M. Bonn, Y. Nagata, Vibrational Spectroscopy and Dynamics of Water. *Chem. Rev.* **116**, 7590–7607 (2016).

25. E. Hand, Cheap Shots. *Science.* **367**, 354–358 (2020).

26. L. V. Bock, H. Grubmüller, Effects of cryo-EM cooling on structural ensembles. *Nat. Commun.* **13**, 1709 (2022).

27. M. G. Pamato, I. G. Wood, D. P. Dobson, S. A. Hunt, L. Vočadlo, The thermal expansion of gold: point defect concentrations and pre-melting in a face-centred cubic metal. *J. Appl. Crystallogr.* **51**, 470–480 (2018).

28. S. Keskin, P. Kunnas, N. de Jonge, Liquid-Phase Electron Microscopy with Controllable Liquid Thickness. *Nano Lett.* **19**, 4608–4613 (2019).

29. Malis, T.; Cheng, S. C.; Egerton, R. F., EELS Log-Ratio Technique for Specimen-Thickness Measurement in the TEM. *J Electron Microsc Tech*. **8**, 193–200 (1988).

30. P. K. Olshin, M. Drabbels, U. J. Lorenz, Characterization of a time-resolved electron microscope with a Schottky field emission gun. *Struct. Dyn.* **7**, 054304 (2020).

31. M. N. Yesibolati, S. Laganá, S. Kadkhodazadeh, E. K. Mikkelsen, H. Sun, T. Kasama, O. Hansen, N. J. Zaluzec, K. Mølhave, Electron inelastic mean free path in water. *Nanoscale*. **12**, 20649–20657 (2020).

32. A. J. Bullen, K. E. O'Hara, D. G. Cahill, O. Monteiro, A. von Keudell, Thermal conductivity of amorphous carbon thin films. *J. Appl. Phys.* **88**, 6317–6320 (2000).

33. J. W. Arblaster, Thermodynamic properties of gold. *J Phase Equilib Diffus*. **37**, 229–245 (2016).

34. J. Huang, Y. Zhang, J. K. Chen, Ultrafast solid–liquid–vapor phase change of a gold film induced by pico- to femtosecond lasers. *Appl. Phys. A*. **95**, 643–653 (2009).

35. W. DeSorbo, W. W. Tyler, The specific heat of graphite from 13° to 300°K. *J. Chem. Phys.* **21**, 1660–1663 (1953).

36. E. Pop, V. Varshney, A. K. Roy, Thermal properties of graphene: Fundamentals and applications. *MRS Bull.* **37**, 1273–1281 (2012).

37. *S.R.P. Silva, Properties of Amorphous Carbon, INSPEC, The Institution of Electrical Engineers, London, 2003.*

38. C. Moelle, M. Werner, F. Szücs, D. Wittorf, M. Sellschopp, J. von Borany, H.-J. Fecht, C. Johnston, Specific heat of single-, poly- and nanocrystalline diamond. *Diam. Relat. Mater.* **7**, 499–503 (1998).

39. K. Murata, H. Tanaka, Liquid–liquid transition without macroscopic phase separation in a water– glycerol mixture. *Nat. Mater.* **11**, 436–443 (2012).

40. C. A. Angell, W. J. Sichina, M. Oguni, Heat capacity of water at extremes of supercooling and superheating. *J. Phys. Chem.* **86**, 998–1002 (1982).

41. H. Pathak, A. Späh, N. Esmaeildoost, J. A. Sellberg, K. H. Kim, F. Perakis, K. Amann-Winkel, M. Ladd-Parada, J. Koliyadu, T. J. Lane, C. Yang, H. T. Lemke, A. R. Oggenfuss, P. J. M. Johnson, Y. Deng, S. Zerdane, R. Mankowsky, P. Beaud, A. Nilsson, Enhancement and maximum in the isobaric specific-heat capacity measurements of deeply supercooled water using ultrafast calorimetry. *Proc. Natl. Acad. Sci.* **118**, e2018379118 (2021).

42. Eric W. Lemmon, Ian H. Bell, Marcia L. Huber, and Mark O. McLinden, "Thermophysical Properties of Fluid Systems" in *NIST Chemistry WeBook, NIST Standard Reference Database Number 69* (National Institute of Standards and Technology, Gaithersburg, MD).

43. L. M. Shulman, The heat capacity of water ice in interstellar or interplanetaryconditions. *Astron. Astrophys.* **416**, 187–190 (2004).

44. O. Andersson, H. Suga, Thermal conductivity of low-density amorphous ice. *Solid State Commun*. **91**, 985–988 (1994).

45. S. J. Byrnes, Multilayer optical calculations. *ArXiv160302720 Phys.* (2020) (available at http://arxiv.org/abs/1603.02720).

46. V. Kofman, J. He, I. Loes ten Kate, H. Linnartz, The Refractive Index of Amorphous and Crystalline Water Ice in the UV–vis. *Astrophys. J.* **875**, 131 (2019).

47. X. Wang, Y. P. Chen, D. D. Nolte, Strong anomalous optical dispersion of graphene: complex refractive index measured by Picometrology. *Opt. Express*. **16**, 22105 (2008).

48. G. Rosenblatt, B. Simkhovich, G. Bartal, M. Orenstein, Nonmodal Plasmonics: Controlling the Forced Optical Response of Nanostructures. *Phys. Rev. X*. **10**, 011071 (2020).

49. W. W. Duley, Refractive indices for amorphous carbon. *Astrophys. J.* **287**, 694 (1984).

50. M. Fuchs, E. Dayan, E. Presnov, Evaporative cooling of a ventilated greenhouse rose crop. *Agric. For. Meteorol.* **138**, 203–215 (2006).

51. J. D. Smith, C. D. Cappa, W. S. Drisdell, R. C. Cohen, R. J. Saykally, Raman thermometry measurements of free evaporation from liquid water droplets. *J Am Chem Soc*. **128**, 12892–12898 (2006).

52. B. E. Warren, *X-ray Diffraction* (Dover Publications, 1990).

53. D. T. Bowron, J. L. Finney, A. Hallbrucker, I. Kohl, T. Loerting, E. Mayer, A. K. Soper, The local and intermediate range structures of the five amorphous ices at 80K and ambient pressure: A Faber-Ziman and Bhatia-Thornton analysis. *J. Chem. Phys.* **125**, 194502 (2006).

54. T. Grant, N. Grigorieff, Automatic estimation and correction of anisotropic magnification distortion in electron microscopes. *J. Struct. Biol.* **192**, 204–208 (2015).

55. R. J. List, *Smithsonian Meteorological Tables* (Smithsonian Institution Press, Washington, 1968).

**Acknowledgments:**

The authors would like to thank Dr. Pavel K. Olshin for his help with the heat transfer simulations as well as Dr. Jonathan M. Voss for his help with the preparation of Fig. 1.

**Funding:**

This work was supported by the ERC Starting Grant 759145 and by the Swiss National Science Foundation Grant PP00P2_163681.

**Author contributions:**

Conceptualization: UJL

Methodology: CRK, NJM, GB, UJL

Investigation: CRK, NJM, GB, UJL

Visualization: CRK, NJM, GB, UJL

Funding acquisition: UJL

Project administration: UJL

Supervision: UJL

Writing – original draft: CRK, NJM, GB, UJL

Writing – review & editing: CRK, NJM, GB, MD, UJL

**Competing interests:** Authors declare that they have no competing interests.

**Data and materials availability:** Positions of the diffraction maxima of water are included in the Supplementary Materials in Data S1. All other data supporting the findings of the study are available from the corresponding author upon reasonable request.

**Supplementary Materials**

Materials and Methods

Supplementary Text

Figs. S1 to S9

Tables S1

Data S1

References (*26–55*)

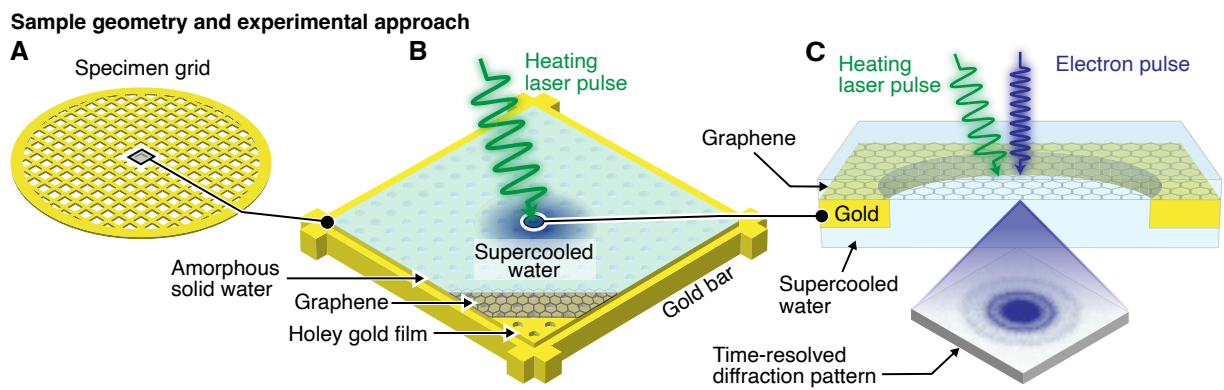

**Fig. 1. Illustration of the experimental approach.** (**A,B**) Illustration of the sample geometry. A gold mesh supports a holey gold film that is covered with few-layer graphene. A 176 nm thick layer of amorphous solid water is deposited (101 K sample temperature), which is then locally heated with a shaped microsecond laser pulse to prepare water in no man's land. (**C**) A diffraction pattern of the supercooled liquid is captured with an intense, 6 µs electron pulse.

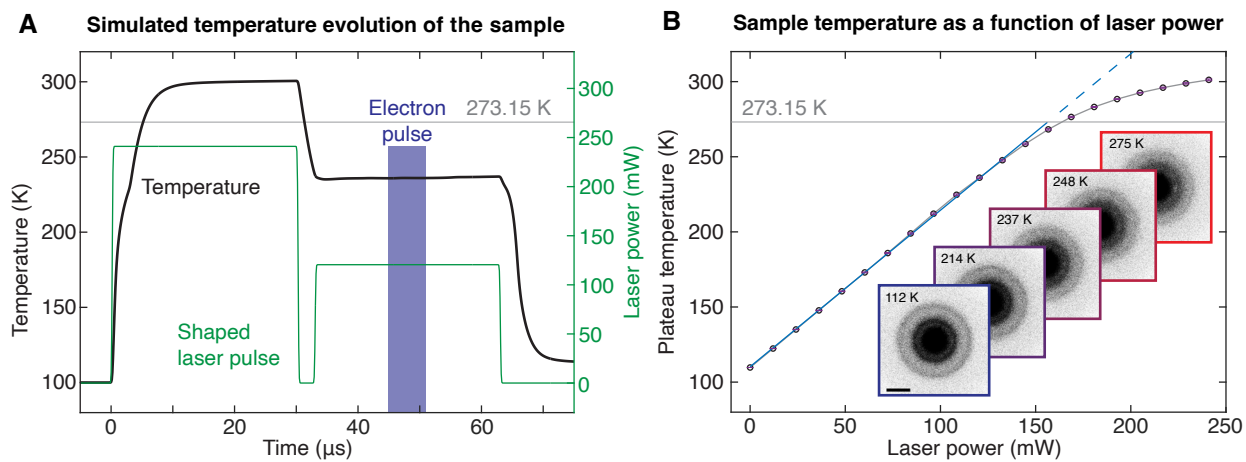

**Fig. 2. Simulation of temperature evolution of the sample.** (**A**) Simulation of the temperature evolution of the sample (black) under irradiation with a shaped microsecond laser pulse (green). The sample is first heated above the melting point and then rapidly cooled to the desired temperature in no man's land by reducing the laser power. Once the temperature has stabilized, we capture a diffraction pattern with a 6 μs electron pulse (blue). (**B**) Simulations show that this temperature increases linearly with laser power. It only starts to level off above ~260 K, where evaporative cooling becomes important. The inset shows diffraction patterns recorded over a range of temperatures. Scale bar, 2 Å⁻¹.

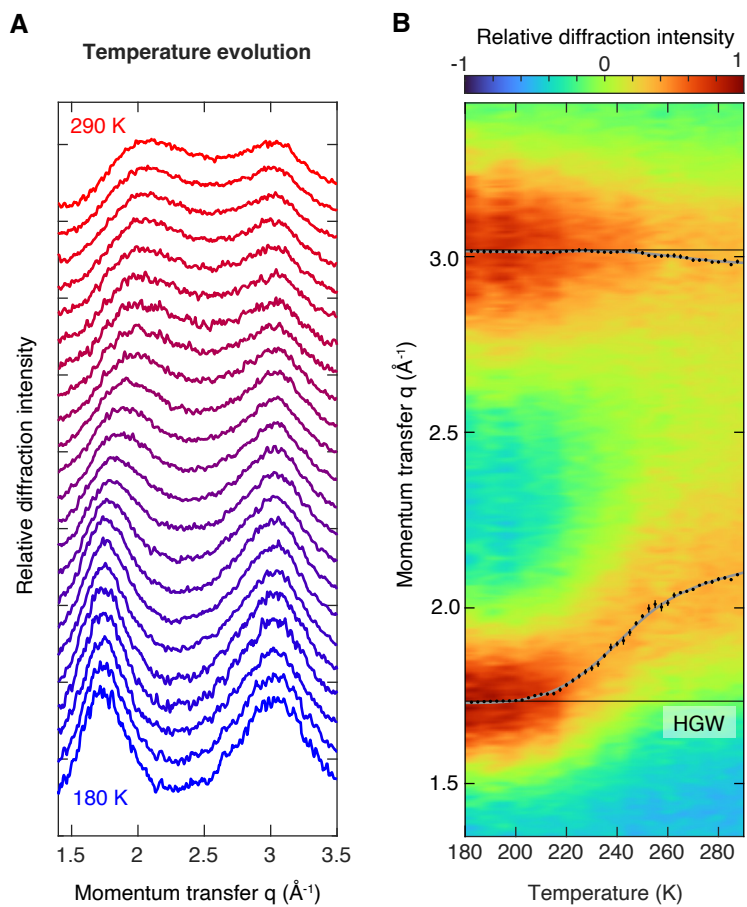

**Fig. 3. Structural evolution of water in no man's land.** (**A**) Diffraction patterns reveal that the structure of water evolves smoothly as it is cooled from 290 K to 180 K. Diffraction patterns are shown in 5 K steps. (**B**) Two-dimensional plot of the evolution of the diffraction pattern of water. Black dots mark the positions of the first two diffraction maxima, with the error bars indicating to the standard error of the mean of five measurements. The grey lines provide a guide to the eye and are derived from splines. The horizontal black lines indicate the positions of the diffraction maxima of HGW.

# Supplementary Materials for

## Electron Diffraction of Water in No Man's Land

Constantin R. Krüger,[†] Nathan J. Mowry,[†] Gabriele Bongiovanni,[†] Marcel Drabbels, and Ulrich J. Lorenz

Correspondence to: ulrich.lorenz@epfl.ch

[†] These authors contributed equally.

**This PDF file includes:**

## Materials and Methods

<u>A. Instrumentation</u>

Experiments were performed with a modified JEOL 2010F transmission electron microscope (Fig. S1A) that we have previously described (*13, 14*). Water in no man's land is prepared *in situ* by melting a sample of amorphous solid water (ASW) on a holey gold film with a shaped microsecond laser pulse (532 nm), which is obtained by modulating the output of a continuous laser with an acousto-optic modulator. The laser beam is directed at the sample with a mirror mounted above the upper pole piece of the objective lens, so that it strikes the sample at close to normal incidence. The laser beam is focused to a spot size of 38 μm FWHM in the sample plane, as determined from an image of the laser beam recorded with a CCD camera that is placed in a conjugate plane.

Figure S1B displays the shape of a typical laser pulse that we use to prepare water in no man's land, as recorded with a fast photodiode. A 30 μs rectangular pulse is used to heat the sample of ASW to a temperature of over 300 K, after which the laser power is reduced, here to 40 %, in order to rapidly supercool the liquid to a well-defined temperature. The rise and fall times of the laser pulse are 250 ns, as determined from Fig. S1B. Note that in order to achieve faster cooling, we initially reduce the laser power by twice the desired amount for a duration of 3 μs. For final laser powers of 50 % or less, we switch the laser off entirely for the same duration. At a delay of 15 μs after reducing the laser power, when the temperature of the liquid has stabilized, we capture a time-resolved diffraction pattern with an intense, high-brightness electron pulse of 6 μs duration as indicated with a blue rectangle in Fig. S1B. We generate such electron pulses as previously described, by temporarily boosting the emission from the Schottky emitter of our microscope to near its limit (*13, 14*). To this end, we briefly heat the emitter tip to extreme temperatures through irradiation with a microsecond laser pulse (532 nm, 1.5 W, 17 μm FWHM spot size at the tip, 150 μs pulse duration, as obtained by chopping the output of a continuous laser with an acousto-optic modulator), which causes the emission current to increase by up to 3.7 times. An electrostatic deflector is then used to slice a 6 μs electron pulse out of the boosted electron beam. As illustrated in Fig. 1A-C, time-resolved electron diffraction patterns are collected from within the central hole of a grid square, with the electron beam converged to a disk of about 1.5 μm diameter. The diffraction patterns are recorded with a TVIPS XF416 electron camera. The camera length is calibrated with the diffraction pattern of the polycrystalline holey gold film (*27*).

<u>B. Fabrication of sample supports</u>

Sample supports are fabricated with the process illustrated in Fig. S2. A holey gold film is prepared by vapor depositing 50 nm of gold onto a holey carbon specimen grid (QUANTIFOIL N1-C15nCu20-01) which consists of a 12 nm holey carbon film (2 μm diameter holes with 1 μm spacing) on a 200 mesh copper grid (Fig. S2A,B). The copper mesh is then etched away by floating the grid on an ammonium persulfate solution, until only the holey thin film is left (Fig. S2C). The remaining etchant is removed in three washing steps, each consisting in transferring the thin film to a fresh bath of deionized water for approximately 10 min. The holey thin film is then transferred onto a 600 mesh gold grid (Plano, 13.5 μm bar width and 8.75 μm bar height) by submerging the grid into the water bath and using it to gently pick up the film (Fig. S2D). The assembly is placed onto a hot plate (50 °C) for up to 2 hours to evaporate any remaining water. In the final step, multilayer graphene is transferred onto the grid assembly. To this end, 6-8 layer graphene on copper foil (Graphene Supermarket) is floated on an ammonium persulfate solution until the copper has dissolved. The multilayer graphene film, which remains floating on the etchant, is then washed three times by transferring it to a fresh bath of distilled water for 10 min each. Finally, the graphene layer is transferred onto the holey gold film by gently picking it up with the specimen grid (Fig. S2E). Before use, the completed assembly is cleaned for 40 seconds in a hydrogen plasma (Pelco Easiglow, negative polarity, 20 mA current, 0.5 mbar).

## C. Deposition of ASW

The sample support is loaded into the microscope with a single tilt cryo specimen holder (Gatan 914), which is then filled with liquid nitrogen to cool the sample to a temperature of $101 \pm 1$ K, which is continuously monitored during the experiment. Amorphous solid water is deposited by leaking water vapor into the column of the microscope through a leak valve that we installed on our instrument for this purpose. Deionized water (15 M$\Omega$ cm) is placed in a stainless-steel reservoir that is water cooled to a temperature of 290 K. After filling the reservoir, the water is degassed by evacuating the reservoir for 5 minutes with a membrane pump. The water vapor in the reservoir is then leaked into the microscope through a gas dosing valve (Balzers UDV235) that is connected to a 60 cm long stainless-steel tube. The end of the tube protrudes into the cold shield, a small metal enclosure between the pole pieces of the objective lens that surrounds the sample, which is ordinarily cooled to liquid nitrogen temperature to serve as a cryo pump, but in our experiments is held at ambient temperature. The nozzle of the tube has an inner diameter of 1 mm and terminates at approximately 10 mm from the edge of the sample holder. Since the nozzle lies in the sample plane, no direct line of sight exists between the tube and the sample. Water molecules must therefore first undergo collisions with surfaces before they can reach the specimen grid. Accordingly, we find that the deposition rate does not vary across the specimen support. The sample region is pumped through openings in the cold shield, with the surrounding volume of the microscope column evacuated to a pressure of $2 \cdot 10^{-7}$ mbar by a 150 L ion pump when no water is being leaked into the microscope. We adjust the deposition rate, typically about 82 nm/min, by monitoring the pressure on the low-pressure side of the leak valve.

The following procedure is used to control the thickness of the ASW layer that is freshly deposited for each experiment. Before beginning an experimental run, the deposition rate is determined by recording diffraction patterns as a function of deposition time, which are then analyzed as described in section G (Analysis of the diffraction patterns). As shown in Fig. S3, the relative intensity of the diffraction pattern initially increases linearly with the deposition time, but then goes through a maximum as the sample grows thicker and multiple scattering as well as inelastic interactions become more important. By comparing this curve to a reference, we can deduce the deposition rate (typically 82 nm/min) and thus adjust the deposition time (typically around 2 min) to obtain a sample thickness of 176 nm (blue arrow in Fig. S3) with a variation in thickness of approximately 12%. Experiments are only started once the deposition rate has stabilized. The deposition rate is remeasured approximately every 2 h, and the deposition time is adjusted accordingly.

The absolute sample thickness was determined with the log-ratio method (*28*, *29*). A calibration sample was grown to a thickness of 299 nm (black arrow in Fig. S3) and transferred to a JEOL 2200FS transmission electron microscope, which is equipped with an energy filter, allowing us to record energy loss spectra of the ASW sample (*30*). The sample thickness $d$ is obtained from the ratio of the total electron beam intensity $I_t$ to the intensity of the zero loss peak $I_0$,

$$d = \lambda \cdot \ln \frac{I_t}{I_0}.$$

Here, $\lambda$ is the electron inelastic mean free path, which can be calculated according to

$$\lambda \approx \frac{106 \, F \, E_0}{E_m \ln\left(\frac{2\beta E_0}{E_m}\right)},$$

where $F = 0.618$ is a relativistic factor, $E_0$ the accelerating voltage in keV, $\beta = 10$ the collection angle in mrad, and $E_m = 7.12$ the average energy loss in eV (*29*). The value for $E_m$ was interpolated from recent measurements on thin liquid water films (*31*). We obtain a thickness of 299 nm for the calibration sample, which yields a thickness of 176 nm for the samples used in our experiments.

D. Time-resolved diffraction experiments

The following procedure is adopted to establish suitable experimental parameters. As illustrated in Fig. 1A-C, time-resolved electron diffraction patterns are collected from within the central hole of a grid square, with the electron beam converged to a disk of about 1.5 µm diameter. Care is taken to choose a sample area without any tears in the graphene film. The laser beam is centered onto the same hole in the gold film using the method previously described (*16*). Once the deposition rate has been established (section C), a 176 nm thick layer of ASW is deposited, and the approximate minimum laser power is determined (100 mW), with which this layer can be successfully melted and revitrified with a 30 µs laser pulse (*16*). The sample is then irradiated with a 1 s laser pulse of approximately 85 mW power to evaporate any ice within the grid square under observation. Finally, the sample is irradiated with another 3 s laser pulse from a second laser (405 nm, 150 mW, and a beam diameter of approximately 70 µm in the sample plane, centered on the area under observation) in order to also evaporate the ice in adjacent grid squares.

Once the experimental parameters have been established, time-resolved experiments are performed with an automated procedure. A fresh layer of ASW is deposited, and a diffraction pattern is recorded (20 boosted electron pulses of 6 µs duration, fired at 10 Hz repetition rate) which is later used to accurately determine the camera length in each experiment (section G) and normalize the intensity of the time-resolved diffraction pattern. The sample is then irradiated with a shaped microsecond laser pulse in order to prepare water in no man's land, and its structure is probed by capturing a diffraction pattern with a boosted electron pulse of 6 µs duration (section A). Occasionally, diffraction patterns are obtained that feature diffraction spots, indicating that the sample has partially crystallized. In this case, the diffraction pattern is discarded, and the experiment is repeated. If crystallization occurs again, the experiment is repeated once more with the electron pulse duration reduced to 3 µs. Out of the 148 diffraction patterns included in our analysis, 10 were recorded with a 3 µs electron pulse. Note that we can identify even faint diffraction spots by comparing with a static diffraction pattern that we record after the experiment (20 boosted electron pulses of 6 µs duration, fired at 10 Hz repetition rate). Crystallites that give rise to diffraction spots usually grow in size after the time-resolved diffraction pattern has been recorded, so that the corresponding diffraction features become very prominent in the static diffraction pattern. This makes the identification of crystalline features in the time-resolved diffraction patterns straightforward. Finally, the ice in the grid square under observation as well as in adjacent grid squares is evaporated as described above. A fresh layer of ASW is then deposited, and the experiment is repeated.

The evolution of the diffraction pattern of water between cryogenic temperature and room temperature was scanned five times (see Fig. S7). In each scan, the laser power in the second part of the shaped laser pulse, during which the diffraction pattern is captured, was increased in small steps. A total of about 30 steps were recorded per scan. The temperature range between about 200 K and 290 K was sampled more finely, with an average step size of about 4 K. The diffraction patterns were then analyzed as described in section G (Analysis of the diffraction patterns) and a temperature calibration was obtained for each scan as detailed in section F (Determination of sample temperature). The five scans were then averaged to obtain the temperature evolution of the diffraction pattern presented in Fig. 3 and Fig. S9.

### E. Simulation of the temperature evolution of the sample

Finite element heat transfer simulations of the temperature evolution of the sample under illumination with microsecond laser pulses are performed using COMSOL Multiphysics. The simulation geometry is illustrated in Fig. S4. A 600 mesh gold grid (13.5 μm wide and 8.75 μm thick bars, 27 μm × 27 μm viewing area) supports a thin film of 50 nm gold on 12 nm amorphous carbon that features a square pattern of holes (2 μm diameter, 1 μm separation) and is covered with a film of 7-layer graphene (2.415 nm thickness). The holey thin film and graphene are covered on both sides by an 88 nm thick layer of ice, for a total of 176 nm. To reduce the computational cost, the graphene film is omitted outside of a 9 μm × 9 μm square area in the center of the geometry, and the extent of the ice-covered holey gold/carbon film is limited to a square of 125 μm side length. To account for the large heat capacity of the specimen grid, the gold bars of the supporting mesh extend another 42.5 μm beyond this square.

The material parameters used in our simulations are listed in Table S1. We use literature values for the temperature-dependent heat capacity and thermal conductivity of amorphous carbon (*32*), gold (*33, 34*), and graphene (*35, 36*). Since reliable low-temperature values for the heat capacity of amorphous carbon are not available (*37*), we use its room temperature value (*38*) and extrapolate it to low temperatures by assuming that the temperature dependence is the same as for graphite (*35*). Experimental values for the heat capacity and thermal conductivity of bulk supercooled water in no man's land are largely not available (*11, 39*). For the heat capacity of water, we use experimental values between 227 K and 350 K (*40–42*) as well as between 100 K and 136 K (*43*). The heat capacity at intermediate temperatures is obtained from a spline interpolation. For the thermal conductivity of water, we use its room temperature value, which is similar to that of amorphous ice at low temperatures (*44*).

The temperature of the entire sample is initially set to 100 K. We simulate heating with a shaped microsecond laser pulse by placing two Gaussian surface heat sources (38 μm FWHM) in the center of the simulation geometry. In order to account for the different absorption of the graphene covered gold film and of the free-standing graphene areas (46 % and 12.5 % of the incident laser radiation, respectively), one heat source is placed on top of the gold film and one on top of the free-standing graphene. The absorbed fractions were determined with the help of a multilayer thin film transfer matrix calculator (*45*), using the complex refractive indices of amorphous ice (*46*), graphene (*47*), gold (*48*), and amorphous carbon (*49*). The heating rates of the Gaussian surface heat sources are calculated from the absorbed fractions and the incident laser power. The temporal profile of the simulated laser pulse closely mimics that in the experiment (section A). We use error functions to simulate the rise and fall times of 250 ns. To account for evaporative cooling, we place negative heat sources on the top and on the bottom surface of the water film. The cooling rate is determined from the temperature-dependent enthalpy of evaporation (*50*) and the temperature-dependent evaporation rate of water, as calculated with the formula in Table S1 (*51*). The temperature of the ice is probed within the hole of the gold film that is located in the center of the geometry. We report the average temperature and its standard deviation during the electron pulse within a cylindrical volume of 1.5 μm diameter that spans the entire thickness of the ice film and that is centered on the central hole.

F. Determination of the sample temperature

As detailed in section D (Time-resolved diffraction experiments), the evolution of the diffraction pattern of water between cryogenic temperature and room temperature was scanned five times. In order to obtain a temperature calibration for each scan, we make use of the fact that the plateau temperature of the sample during the second part of the shaped laser pulse increases linearly with laser power for temperatures below the melting point (Fig. 2B). This linear dependence can be understood by considering that the temperature is stable when the heating rate, which is proportional to the laser power, equals the rate of heat dissipation. The latter is proportional to the temperature difference between the area under observation and the grid bars, which remain at cryogenic temperature. As a result, the plateau temperature increases linearly with laser power as long as evaporative cooling is negligible. We can therefore calibrate the scan in this linear range by determining the plateau temperature at two different laser powers and interpolating linearly — at zero laser power (*i.e.* the laser is switched off entirely after 30 μs) as well as at high laser power. At zero laser power, the temperature is close to that of the sample before the laser pulse, with a small offset that we determine from simulations. The high temperature calibration point is obtained by comparison with x-ray data. As described below, we reduce the error of our calibration by also including data points above the melting point, where evaporative cooling causes deviations from linearity (Fig. 2B). We are able to do so by correcting for the effect of evaporative cooling, which effectively reduces the laser power available to heat the sample.

Simulations show that the sample temperature at zero laser power is slightly higher than 100 K, the temperature of the sample before the laser pulse. This is due to the finite heat capacity of the sample support, which slightly warms up under laser irradiation. We determine the sample temperature at zero laser power from our heat transfer simulations. As shown in Fig. S5, this temperature (~110 K) slightly increases with the temperature that the sample reaches at maximum laser power. We therefore determine the temperature at zero laser power iteratively. Using an initial guess for the plateau temperature at zero laser power, we obtain a temperature calibration as described below, which yields the temperature at maximum laser power. We then update the temperature at zero laser power accordingly, using the data from Fig. S5, and repeat until the solution converges. Note that any systematic error in the determination of the sample temperature at zero laser power that might potentially arise from discrepancies between simulation and experiment leads to a much smaller error at higher temperatures. For example, one can estimate that the error at 230 K is about 5 times smaller than the error of the plateau temperature at zero laser power.

The temperature of the sample at high laser power is determined by comparing the position of the first diffraction maximum in our experiment with x-ray diffraction data for bulk water at ambient temperature and in the mildly supercooled regime, which were obtained by cooling a water sample in a cryostat (green squares in Fig. S9, (*11*)). We do not use the position of the second diffraction maximum in our calibration since it is less sensitive to temperature changes. Moreover, it contains a small contribution from hydrocarbon contaminations on the graphene support, which appears to affect the peak position at high temperatures.

Note that electron and x-ray diffraction yield small differences in the positions of the diffraction maxima, which largely arise from differences in the diffraction background that is obtained with either method (*52*). In order to make it possible to compare the electron and x-ray diffraction data, the momentum transfer axes are adjusted linearly as shown in Fig. S9, so that the peak positions of amorphous ice in both experiments coincide (horizontal black lines in Fig. S9). For the x-ray data, we use the peak positions of low density amorphous ice (*53*), which are compared to the peak positions of HGW in our experiment.

The following procedure is used to correct for the effect of evaporative cooling, which sets in at elevated temperatures, causing the plateau temperature of the sample to rise more slowly with laser power and to deviate from linearity (Fig. 2B). Evaporative cooling effectively reduces the laser power $P_{laser}$ that is available to heat the sample. The effective heating power $P_{heat}$ that is available therefore becomes

$$P_{heat} = P_{laser} - a \cdot P_{cool}(T)$$

where $P_{cool}(T)$ is the effective power of evaporative cooling, which we determine from our simulations. The relationship between the laser power $P_{laser}$, the power of evaporative cooling $P_{cool}(T)$, and the effective heating power $P_{heat}$ is illustrated in Fig. S6. The fit parameter $a$ accounts for discrepancies between the simulations and the experiment with respect to the absolute laser power that is required to reach a given plateau temperature. These discrepancies likely arise from differences between the actual and simulated absorption coefficients and heat conductivities

The plateau temperature of the sample $T_{sample}$ depends linearly on the effectively available heating power $P_{heat}$ (Fig. S6), so that

$$T_{sample} = T_0 + b \cdot P_{heat}$$

where $T_0$ is the plateau temperature of the sample at zero laser power, which we determine as described above, and $b$ is a fit parameter. We can therefore obtain a calibration of the sample temperature $T_{sample}$ by varying the parameters $a$ and $b$ to optimize the agreement (least squares fitting of the position of the first diffraction maximum) between our electron diffraction data and the x-ray diffraction data for bulk water at high temperatures (green squares in Fig. S9, (*11*)). We estimate the errors of the temperature calibration from the standard errors of the fit parameters using error propagation. Figure S7A shows the positions of the diffraction maxima for each scan after temperature calibration. The errors of the temperature calibration are shown in Fig. S7B.

We note that while the inclusion of data points at high temperature reduces the temperature error, it only nominally affects the calibration. We obtain virtually the same calibration if we only use the three x-ray data points with the lowest temperature (251 K, 253 K, and 258 K, where evaporative cooling is small). The same is true if we additionally neglect the evaporation correction entirely.

G. Analysis of the diffraction patterns

Both time-resolved and static diffraction patterns are analyzed with the following procedure. We first determine the center of each diffraction pattern from the center of mass of the direct beam and then correct for the ellipticity of the diffraction pattern (typically about 1 %). In each experimental run, the distortion is determined from a diffraction pattern of the polycrystalline holey gold film of the sample support, which we record under identical conditions. The distortion parameters are extracted in analogy to a frequently used method for determining the magnification distortion of electron micrographs from their diffractograms (*54*). We then azimuthally average the diffraction patterns and subtract the diffraction background, which consists of the atomic scattering, contributions from inelastic and multiple scattering, as well as the instrument background (*52*). The diffraction background is determined from a logarithmic spline of the diffraction intensity. For each time-resolved diffraction pattern, the camera length is determined from the static diffraction pattern of the ASW sample that we record before the time-resolved experiment. In order to account for variations in sample thickness, we normalize the intensity of the time-resolved diffraction pattern on the intensity of this static diffraction pattern, which typically varies by 5 %. When recorded with 3 μs electron pulses, the intensity of this static diffraction pattern varies by 11%.

The positions of the first two diffraction maxima indicated in Fig. 3 are determined from polynomial fits. For temperatures below ~200 K, the first two diffraction maxima are well separated, allowing us to fit them individually with a 7th order polynomial. At higher temperatures, we fit both maxima together, using an 8th order polynomial. The errors of the positions of the diffraction maxima reported in Fig. S7 are obtained from the standard errors of the fit parameters through error propagation. The error bars shown in Fig. 3 and Fig. S9 represent the standard error of the average peak position of the five scans recorded.

The inflection point of the temperature evolution of the first diffraction maximum (243±2 K) was obtained by calculating the numerical derivative of a smoothing spline of the data in Fig. 3. The standard error was estimated from the scatter of the values obtained for each of the five scans of the temperature evolution.

**Supplementary Text**

<u>H. Measurement of the cooling time</u>

As illustrated in Fig. 2A, we prepare supercooled water by irradiating a sample of ASW with a shaped microsecond laser. The sample is initially heated to room temperature with a 30 µs laser pulse, after which the laser power is reduced to rapidly supercool the liquid. At a time delay of 15 µs, we then probe the structure of the liquid with a 6 µs electron pulse. Time-resolved measurements confirm that at this point in time, the sample temperature has stabilized, allowing us to probe the structure at one unique temperature. Figure S8A shows the evolution of the diffraction pattern of water that ensues when the heating laser is switched off entirely after 30 µs. As the liquid rapidly cools to a temperature of about 110 K, we probe its structural evolution with 2 µs electron pulses. The positions of the first two diffraction maxima are indicated with black dots. The error bars indicate the uncertainty of the fit and are obtained from the standard errors of the fit parameters through error propagation (section G). For comparison, the peak positions of HGW are indicated with horizontal lines. The experiment reveals that the liquid cools so rapidly that at about 40 µs, the structure of water has approached that of HGW, after which it does not change significantly anymore. In the experiment of Fig. 3, the structure of water is probed later, at 45 µs. The timing of the 6 µs electron pulse is indicated with two vertical lines.

In the experiment of Fig. S8B, the laser power is reduced to 40 %, instead of switching it off entirely, so that the sample cools to a temperature of about 235 K. As described in section A (Instrumentation), we initially reduce the laser power to zero for a duration of 3 µs, so as to achieve faster cooling. Fig. S8B reveals that during the time window of the 6 µs electron pulse in the experiment in Fig. 3, the sample temperature has stabilized. We determine a 1/e cooling time of 11 µs in good agreement with simulations. Note that our simulations show that the longest cooling time occurs for plateau temperatures near 228 K, where the heat capacity of supercooled water has a maximum. Therefore, for most other plateau temperatures, cooling will be faster than in Fig. S8B.

<u>I. Temperature evolution of the diffraction pattern of water from Fig. 3, showing a wider temperature range, and comparison with x-ray data</u>

Figure S9 shows the evolution of the diffraction pattern of water between 110 K and 315 K. For comparison, the peak positions measured previously with x-ray diffraction are also shown. Details are given in the caption.

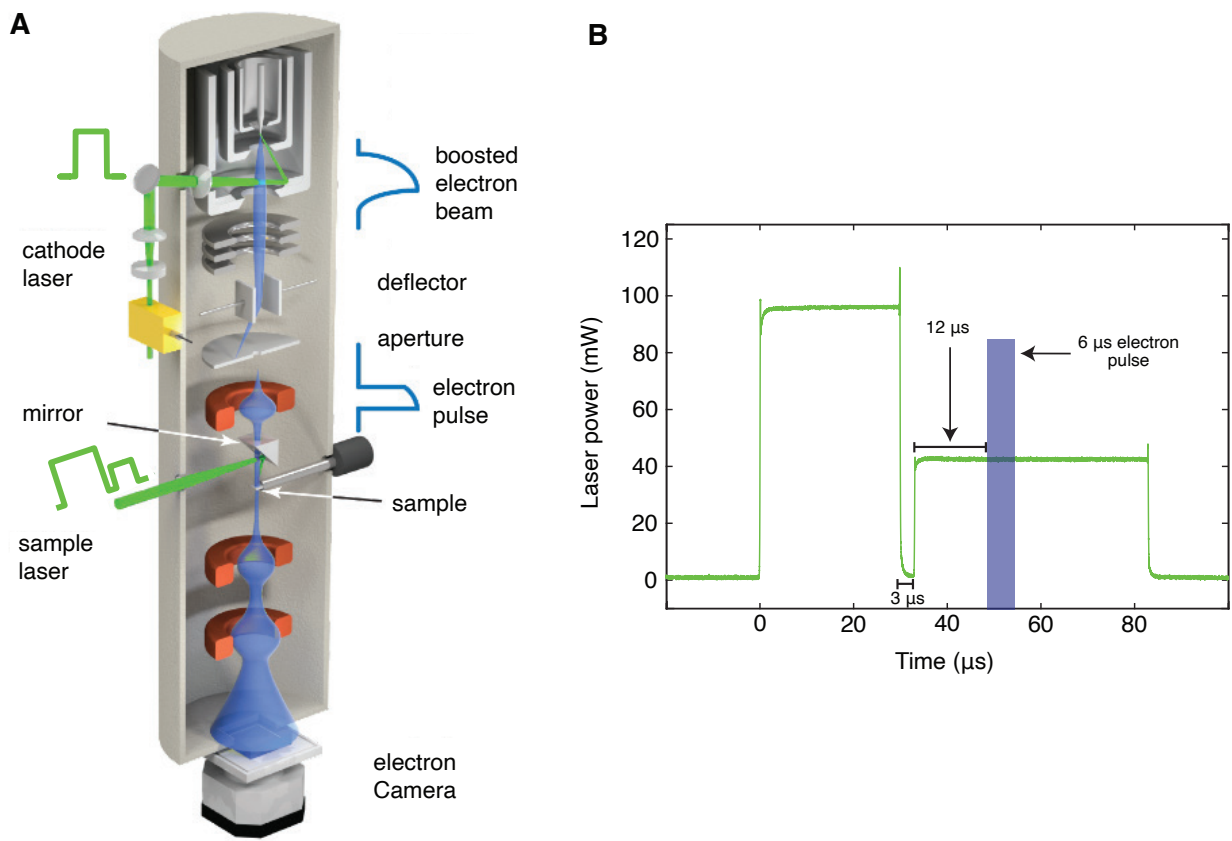

**Fig. S1.**
Illustration of the time-resolved electron microscope and typical shape of the laser pulse used to prepare water in no man's land. (**A**) Illustration of the time-resolved electron microscope. Water in no man's land is prepared *in situ* by melting a sample of ASW with a shaped microsecond laser pulse. Its structure is then probed by recording a time-resolved diffraction pattern with an intense, high-brightness electron pulse of microsecond duration. (**B**) Typical laser pulse shape used to prepare water in no man's land, as recorded with a fast photodiode. The timing of the electron pulse is indicated schematically.

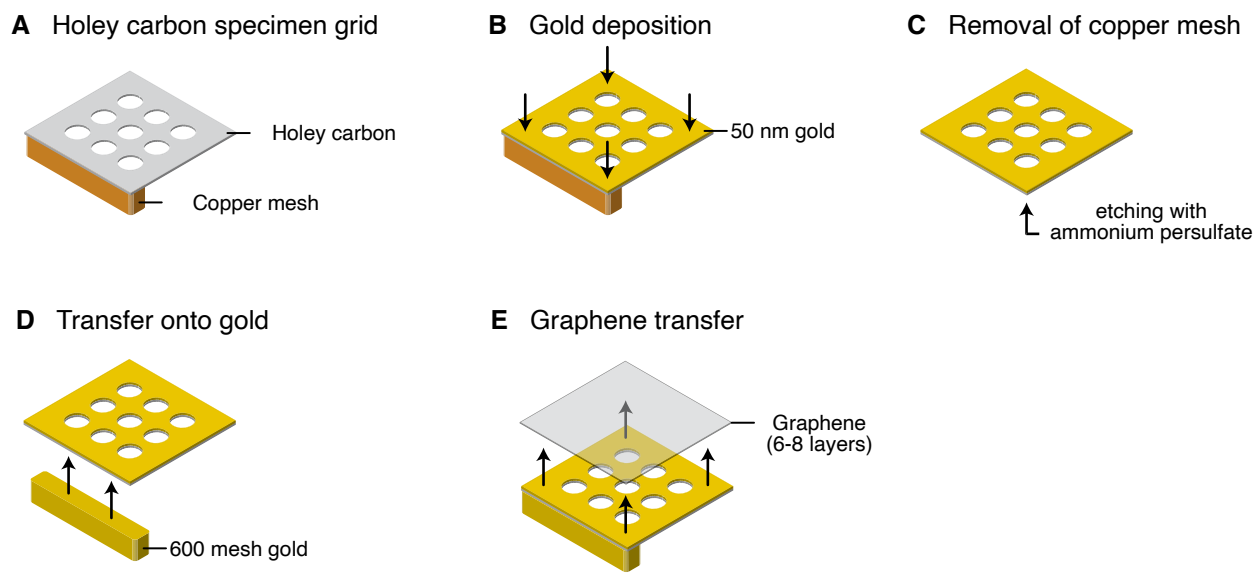

**Fig. S2.**
Fabrication of the sample support. (**A-B**) A holey gold thin film is fabricated by depositing gold onto a holey carbon film on a copper mesh. (**C-E**) After etching away the copper, the holey gold film is picked up with a 600 mesh gold grid, and multilayer graphene is transferred onto the assembly.

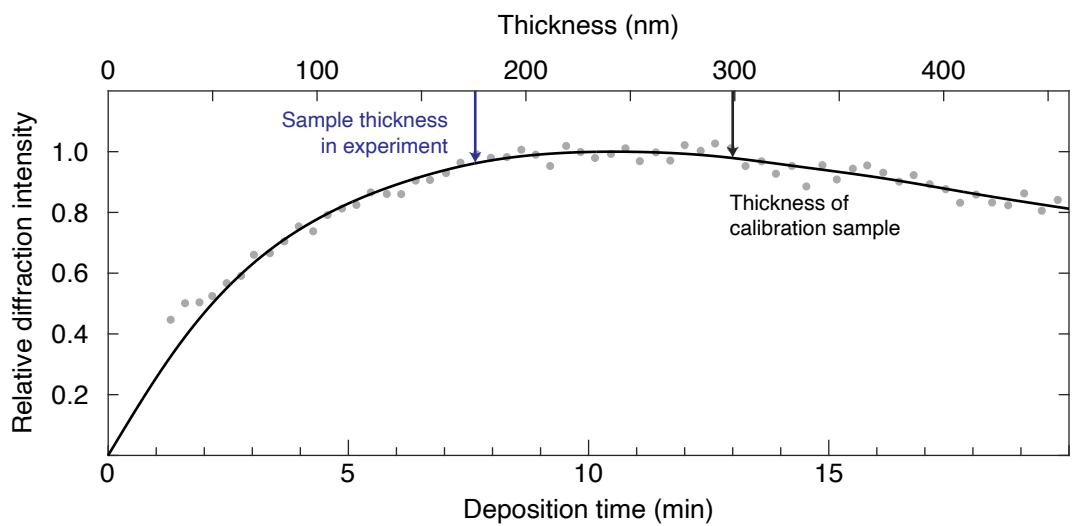

**Fig. S3.**
Relative intensity of the diffraction pattern of water as a function of deposition time. The corresponding sample thickness is indicated on the top axis, as determined from a calibration experiment. The thickness of the calibration sample (299 nm) as well as the sample thickness in our experiment (176 nm) are marked with black and blue arrows, respectively. The solid line is a spline of the data that includes the origin.

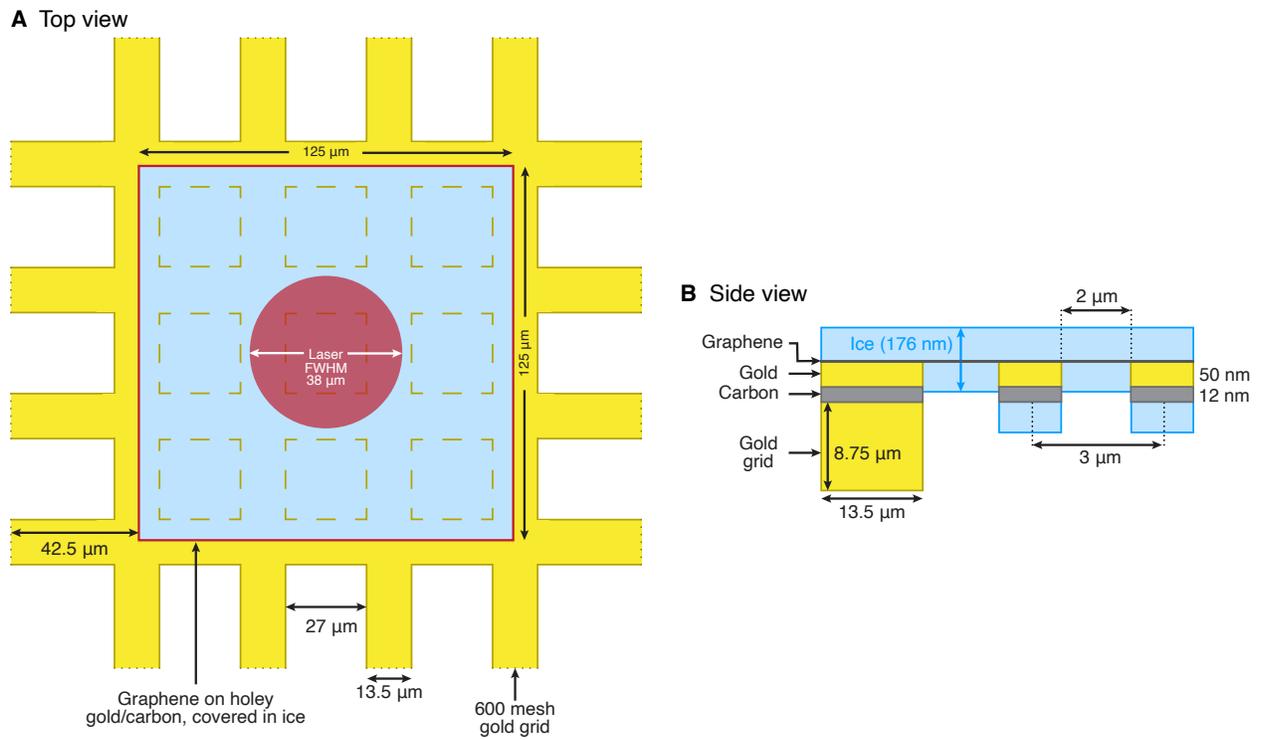

**Fig. S4.**
Sample geometry used in the heat transfer simulations. (**A**) Top view. The laser is centered on the simulation geometry, with the FWHM of the Gaussian laser spot indicated by a red circle. (**B**) Side view.

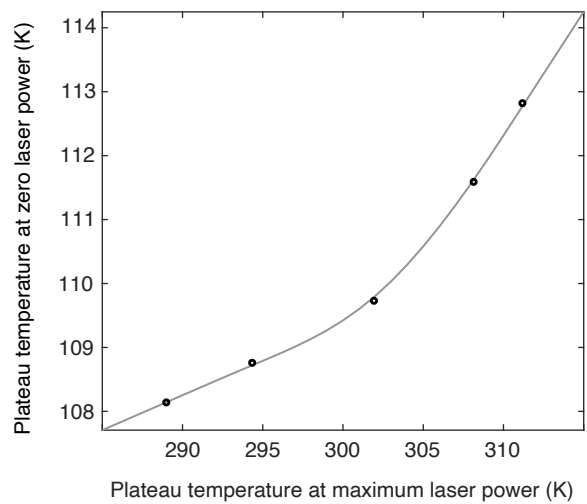

**Fig. S5.**
Simulated plateau temperature at zero laser power as a function of maximum temperature reached.

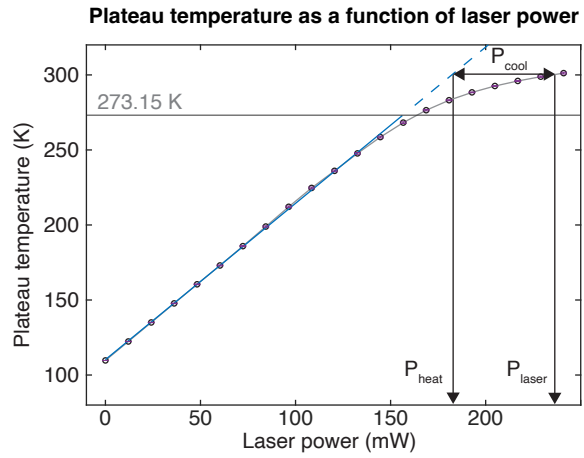

**Fig. S6.**
Simulation of the plateau temperature of the sample as a function of laser power from Fig. 2B and illustration of the relationship between the laser power $P_{laser}$, the power of evaporative cooling $P_{cool}(T)$, and the effective heating power $P_{heat}$. The power that is effectively available to heat the sample $P_{heat}$ is the difference of the laser power $P_{laser}$ and the cooling power $P_{cool}(T)$ that is associated with evaporative cooling.

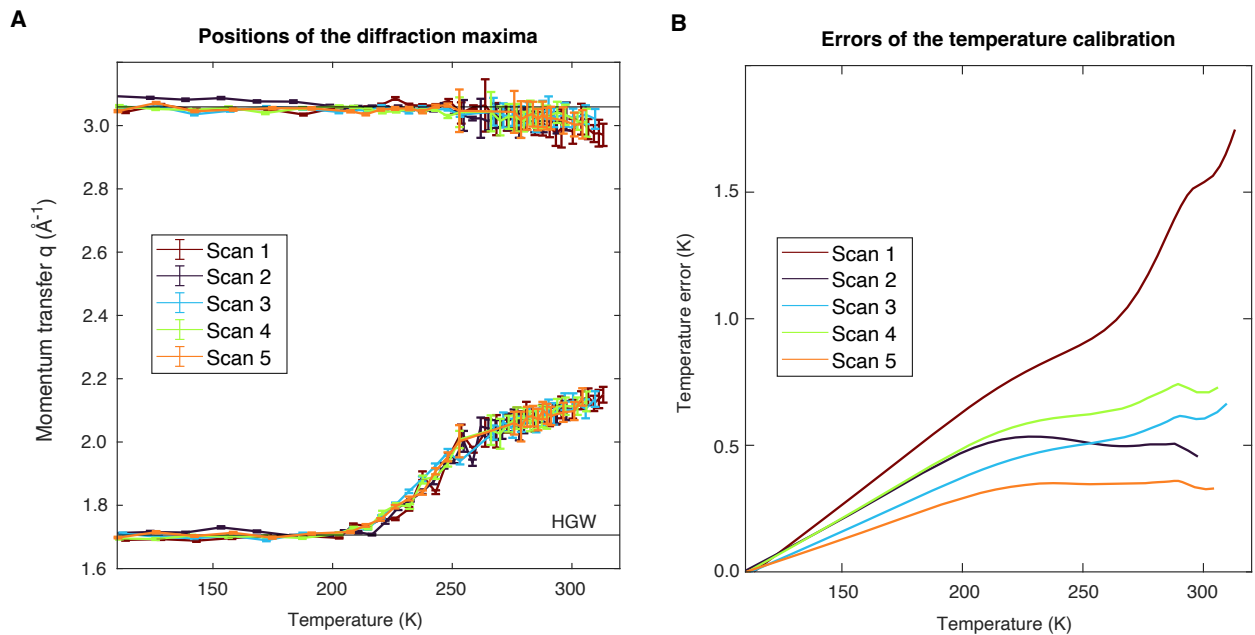

**Fig. S7.**
Positions of the diffraction maxima of water for each scan after temperature calibration as well as errors of the temperature calibration. (**A**) Evolution of the peak positions as a function of sample temperature. The error bars are derived as detailed in section G (Analysis of the diffraction patterns). (**B**) Errors of the temperature calibration.

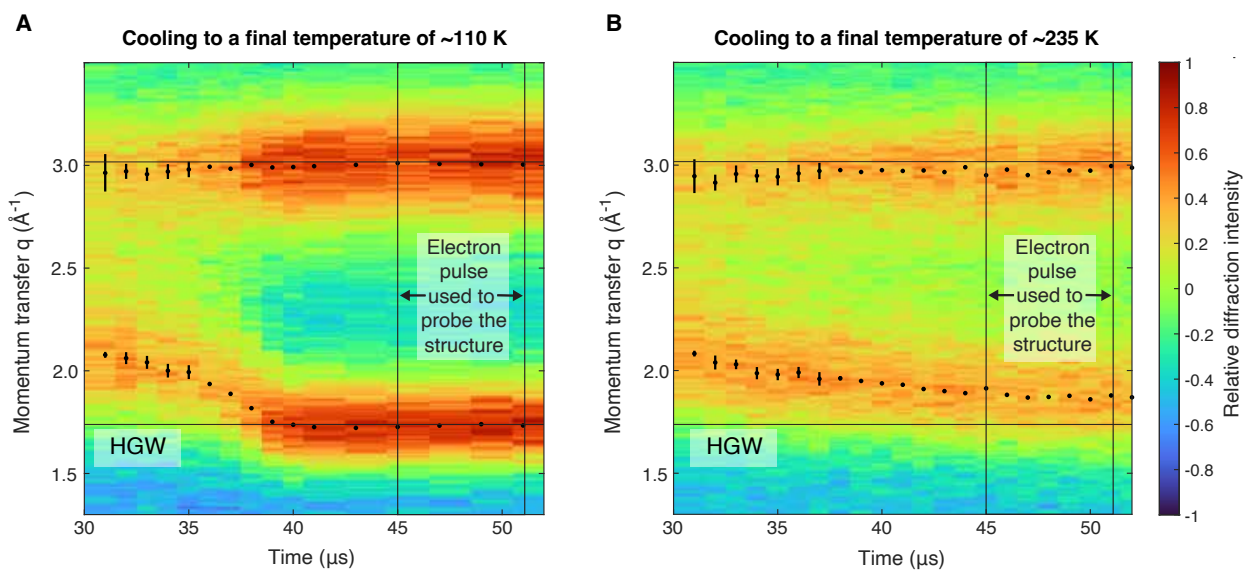

**Fig. S8.**
Temporal evolution of the diffraction pattern of water after the laser power is reduced to rapidly cool the liquid. As shown in Fig. 2A, the sample is initially heated with a 30 µs laser pulse and reaches a temperature close to room temperature, after which the laser power is reduced. As the liquid rapidly cools, we probe its structural evolution with 2 µs electron pulses. (**A**) The laser is switched off entirely, and the sample rapidly cools to a temperature of about 110 K. The positions of the first two diffraction maxima are indicated with black dots, with the error bars determined from the errors of the fit parameters using error propagation. The peak positions of HGW are indicated with horizontal lines. The timing of the 6 µs electron pulses used to probe the structure of water in Fig. 3 is indicated with two vertical lines. (**B**) The laser power is reduced to 40 %, so that the sample cools to a temperature of about 235 K.

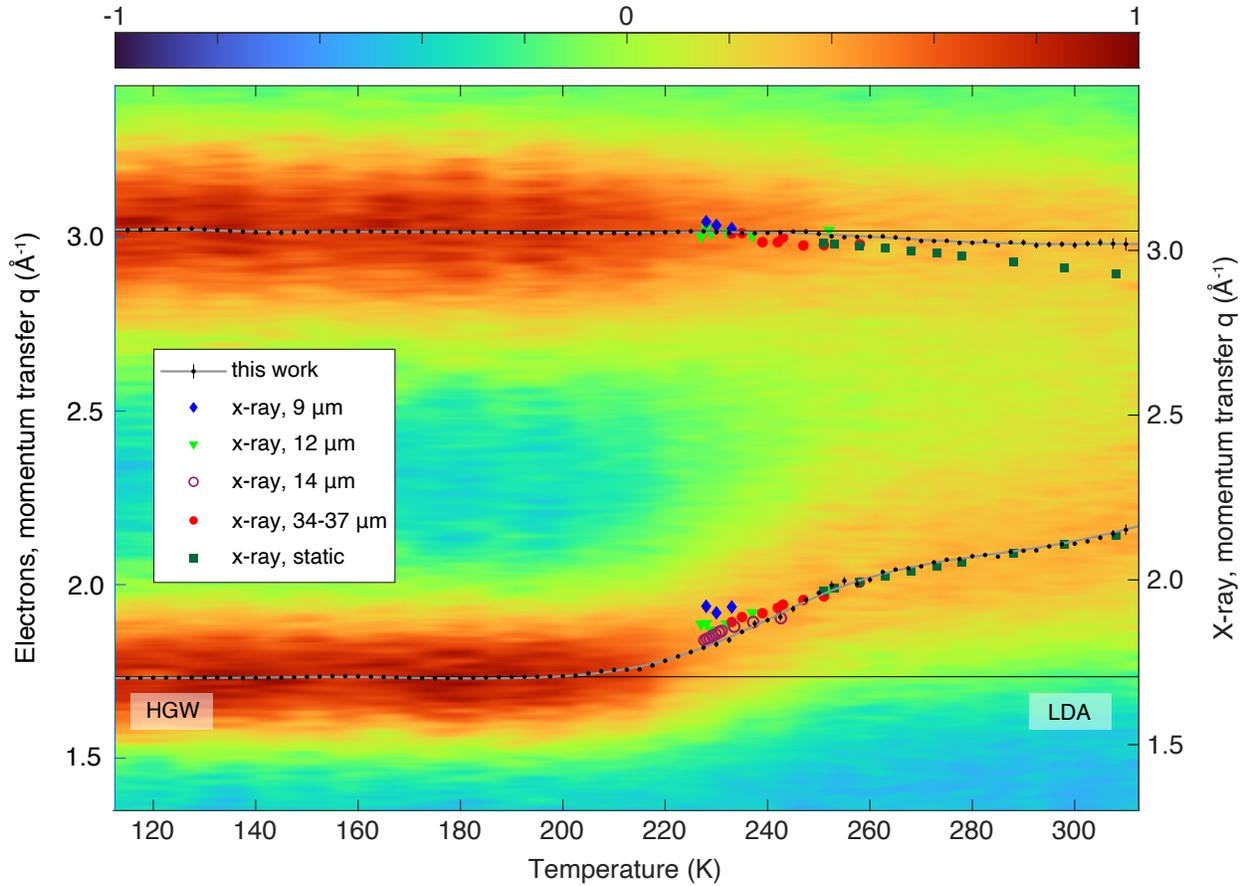

**Fig. S9.**
Evolution of the diffraction pattern of water between 110 K and 315 K. The positions of the first two diffraction maxima are indicated with black dots. The error bars correspond to the standard error of the mean of five measurements. The grey lines provide a guide to the eye and are derived from splines. For comparison, the positions of the diffraction maximum as previously obtained from x-ray diffraction are also shown. Green squares correspond to a measurement in which water was cooled with a cryostat to reach the supercooled regime (*11*), while blue diamonds (*11*), green triangles (*11*), closed circles (*11*) and open red circles (*18*) refer to measurements on evaporatively cooled microdroplets. Note that electron and x-ray diffraction yield small differences in the positions of the diffraction maxima, which largely arise from differences in the diffraction background that is obtained with either method. In order to make it possible to compare the electron and x-ray diffraction data, the momentum transfer axes (electron left, x-rays right) are adjusted linearly such that the peak positions of amorphous ice in both experiments coincide (horizontal black lines). For our electron diffraction data, we use the peak positions of HGW, which we obtain in our experiment when we rapidly cool the sample to cryogenic temperature. For the x-ray data, we use peak positions from Ref. (*53*).

| Property | Value | Ref. |
|---|---|---|
| Heat capacity of gold | $38.5679 + 1.2434 \cdot T - 7.137 \cdot 10^{-3} \cdot T^2 + 1.9237 \cdot 10^{-5} \cdot T^3 - 1.9801 \cdot 10^{-8} \cdot T^4$ (J/kg·K) | ($33$) |
| Thermal conductivity of gold | $320.973 - 0.0111 \cdot T - 2.747 \cdot 10^{-5} \cdot T^2 - 4.048 \cdot 10^{-9} \cdot T^3$ (W/m·K) | ($34$) |
| Heat capacity of graphene | Data from Table 2 of Ref. ($35$) | ($35$) |
| Thermal conductivity of graphene | Data for supported graphene from Fig. 3A of Ref. ($36$) | ($36$) |
| Heat capacity of water | Splined data from Ref. ($40$–$43$) | ($40$–$43$) |
| Thermal conductivity of water | 0.6 W/(m·K) | ($44$) |
| Absorption of thin film (ice – graphene – gold – carbon – ice) | 46 % at 532 nm<br>Refractive indices used in calculating the absorption<br>ice 1.30, graphene 2.67 + i1.34, gold 0.47 + i2.17, carbon 2.28 + i0.63 | ($46$–$49$) |
| Absorption of free-standing graphene (ice – graphene – ice) | 12.5 % at 532 nm for 7 layer graphene (2.415 nm thickness) | ($46$, $47$) |
| Enthalpy of evaporation of water | $2.498 \cdot 10^6 - 3.369 \cdot 10^3 \cdot T$ (J/kg) | ($50$) |
| Evaporation rate | $J_{c,obs} = \gamma_e J_{e,max} = \frac{\gamma P_0}{\sqrt{2\pi m k_B T}}$, where $\gamma_e = 1.0$ (Eq. 2 of Ref. ($51$)) | ($51$) |
| Vapor pressure of water | Equation 1 on page 350 of Ref. ($55$) | ($55$) |

**Table S1.**
Material properties used in the heat transfer simulations.

| Temperature (K) | Position Peak 1 ($Å^{-1}$) | Position Peak 2 ($Å^{-1}$) | Error Peak 1 ($Å^{-1}$) | Error Peak 2 ($Å^{-1}$) |
|---|---|---|---|---|
| 112.5 | 1.7306 | 3.0202 | 0.0041 | 0.0081 |
| 115.0 | 1.7309 | 3.0213 | 0.0040 | 0.0074 |
| 117.5 | 1.7312 | 3.0224 | 0.0040 | 0.0067 |
| 120.0 | 1.7315 | 3.0235 | 0.0041 | 0.0062 |
| 122.5 | 1.7318 | 3.0246 | 0.0043 | 0.0058 |
| 125.0 | 1.7321 | 3.0257 | 0.0046 | 0.0056 |
| 127.5 | 1.7321 | 3.0260 | 0.0045 | 0.0054 |
| 130.0 | 1.7318 | 3.0240 | 0.0042 | 0.0055 |
| 132.5 | 1.7314 | 3.0217 | 0.0040 | 0.0057 |
| 135.0 | 1.7310 | 3.0193 | 0.0039 | 0.0060 |
| 137.5 | 1.7307 | 3.0169 | 0.0039 | 0.0063 |
| 140.0 | 1.7307 | 3.0148 | 0.0042 | 0.0069 |
| 142.5 | 1.7307 | 3.0132 | 0.0047 | 0.0076 |
| 145.0 | 1.7318 | 3.0141 | 0.0050 | 0.0075 |
| 147.5 | 1.7331 | 3.0151 | 0.0052 | 0.0074 |
| 150.0 | 1.7343 | 3.0161 | 0.0055 | 0.0073 |
| 152.5 | 1.7355 | 3.0171 | 0.0058 | 0.0073 |
| 155.0 | 1.7361 | 3.0178 | 0.0056 | 0.0069 |
| 157.5 | 1.7364 | 3.0180 | 0.0051 | 0.0064 |
| 160.0 | 1.7357 | 3.0176 | 0.0047 | 0.0059 |
| 162.5 | 1.7346 | 3.0172 | 0.0043 | 0.0056 |
| 165.0 | 1.7335 | 3.0169 | 0.0040 | 0.0053 |
| 167.5 | 1.7325 | 3.0165 | 0.0038 | 0.0052 |
| 170.0 | 1.7314 | 3.0163 | 0.0037 | 0.0053 |
| 172.5 | 1.7303 | 3.0165 | 0.0036 | 0.0055 |
| 175.0 | 1.7300 | 3.0164 | 0.0029 | 0.0053 |
| 177.5 | 1.7305 | 3.0162 | 0.0020 | 0.0052 |
| 180.0 | 1.7310 | 3.0159 | 0.0013 | 0.0053 |
| 182.5 | 1.7314 | 3.0156 | 0.0011 | 0.0054 |
| 185.0 | 1.7322 | 3.0151 | 0.0016 | 0.0055 |
| 187.5 | 1.7332 | 3.0144 | 0.0023 | 0.0056 |
| 190.0 | 1.7339 | 3.0144 | 0.0021 | 0.0048 |
| 192.5 | 1.7345 | 3.0144 | 0.0021 | 0.0040 |
| 195.0 | 1.7350 | 3.0143 | 0.0021 | 0.0033 |
| 197.5 | 1.7355 | 3.0142 | 0.0023 | 0.0029 |
| 200.0 | 1.7364 | 3.0140 | 0.0025 | 0.0028 |
| 202.5 | 1.7377 | 3.0136 | 0.0030 | 0.0032 |
| 205.0 | 1.7443 | 3.0131 | 0.0009 | 0.0027 |
| 207.5 | 1.7504 | 3.0135 | 0.0027 | 0.0025 |
| 210.0 | 1.7543 | 3.0128 | 0.0036 | 0.0030 |
| 212.5 | 1.7547 | 3.0115 | 0.0039 | 0.0036 |
| 215.0 | 1.7561 | 3.0114 | 0.0051 | 0.0035 |
| 217.5 | 1.7672 | 3.0135 | 0.0061 | 0.0031 |
| 220.0 | 1.7806 | 3.0155 | 0.0059 | 0.0035 |
| 222.5 | 1.7933 | 3.0165 | 0.0060 | 0.0045 |
| 225.0 | 1.8058 | 3.0182 | 0.0069 | 0.0059 |
| 227.5 | 1.8182 | 3.0181 | 0.0076 | 0.0059 |
| 230.0 | 1.8282 | 3.0155 | 0.0081 | 0.0043 |
| 232.5 | 1.8403 | 3.0135 | 0.0086 | 0.0031 |
| 235.0 | 1.8640 | 3.0128 | 0.0063 | 0.0027 |
| 237.5 | 1.8878 | 3.0121 | 0.0103 | 0.0034 |
| 240.0 | 1.8975 | 3.0135 | 0.0069 | 0.0026 |
| 242.5 | 1.9068 | 3.0153 | 0.0116 | 0.0027 |
| 245.0 | 1.9302 | 3.0174 | 0.0096 | 0.0044 |
| 247.5 | 1.9560 | 3.0173 | 0.0064 | 0.0071 |
| 250.0 | 1.9763 | 3.0108 | 0.0060 | 0.0055 |
| 252.5 | 1.9984 | 3.0034 | 0.0117 | 0.0025 |
| 255.0 | 2.0107 | 3.0010 | 0.0113 | 0.0027 |
| 257.5 | 2.0023 | 3.0017 | 0.0122 | 0.0046 |
| 260.0 | 2.0139 | 3.0028 | 0.0084 | 0.0057 |
| 262.5 | 2.0362 | 3.0034 | 0.0075 | 0.0062 |
| 265.0 | 2.0436 | 3.0004 | 0.0043 | 0.0063 |
| 267.5 | 2.0471 | 2.9968 | 0.0014 | 0.0034 |
| 270.0 | 2.0526 | 2.9901 | 0.0041 | 0.0050 |
| 272.5 | 2.0643 | 2.9899 | 0.0016 | 0.0041 |
| 275.0 | 2.0718 | 2.9906 | 0.0045 | 0.0037 |
| 277.5 | 2.0741 | 2.9845 | 0.0058 | 0.0039 |
| 280.0 | 2.0819 | 2.9847 | 0.0060 | 0.0034 |
| 282.5 | 2.0855 | 2.9894 | 0.0045 | 0.0052 |
| 285.0 | 2.0805 | 2.9773 | 0.0016 | 0.0040 |
| 287.5 | 2.0945 | 2.9861 | 0.0030 | 0.0056 |
| 290.0 | 2.0977 | 2.9857 | 0.0046 | 0.0087 |
| 292.5 | 2.0985 | 2.9775 | 0.0029 | 0.0091 |
| 295.0 | 2.1103 | 2.9800 | 0.0045 | 0.0085 |
| 297.5 | 2.1165 | 2.9809 | 0.0055 | 0.0083 |
| 300.0 | 2.1178 | 2.9777 | 0.0049 | 0.0094 |
| 302.5 | 2.1247 | 2.9826 | 0.0013 | 0.0088 |
| 305.0 | 2.1351 | 2.9861 | 0.0067 | 0.0109 |
| 307.5 | 2.1468 | 2.9826 | 0.0104 | 0.0152 |
| 310.0 | 2.1582 | 2.9811 | 0.0151 | 0.0184 |
| 312.5 | 2.1748 | 2.9808 | 0.0186 | 0.0211 |

**Data S1.**
Positions of the first two diffraction maxima of water as a function of temperature and respective errors (standard error of the mean of five measurements) from Fig. 3 and Fig. S9.

## References

26. L. V. Bock, H. Grubmüller, Effects of cryo-EM cooling on structural ensembles. *Nat. Commun.* **13**, 1709 (2022).

27. M. G. Pamato, I. G. Wood, D. P. Dobson, S. A. Hunt, L. Vočadlo, The thermal expansion of gold: point defect concentrations and pre-melting in a face-centred cubic metal. *J. Appl. Crystallogr.* **51**, 470–480 (2018).

28. S. Keskin, P. Kunnas, N. de Jonge, Liquid-Phase Electron Microscopy with Controllable Liquid Thickness. *Nano Lett.* **19**, 4608–4613 (2019).

29. Malis, T.; Cheng, S. C.; Egerton, R. F., EELS Log-Ratio Technique for Specimen-Thickness Measurement in the TEM. *J Electron Microsc Tech*. **8**, 193–200 (1988).

30. P. K. Olshin, M. Drabbels, U. J. Lorenz, Characterization of a time-resolved electron microscope with a Schottky field emission gun. *Struct. Dyn.* **7**, 054304 (2020).

31. M. N. Yesibolati, S. Laganá, S. Kadkhodazadeh, E. K. Mikkelsen, H. Sun, T. Kasama, O. Hansen, N. J. Zaluzec, K. Mølhave, Electron inelastic mean free path in water. *Nanoscale*. **12**, 20649–20657 (2020).

32. A. J. Bullen, K. E. O'Hara, D. G. Cahill, O. Monteiro, A. von Keudell, Thermal conductivity of amorphous carbon thin films. *J. Appl. Phys.* **88**, 6317–6320 (2000).

33. J. W. Arblaster, Thermodynamic properties of gold. *J Phase Equilib Diffus*. **37**, 229–245 (2016).

34. J. Huang, Y. Zhang, J. K. Chen, Ultrafast solid–liquid–vapor phase change of a gold film induced by pico- to femtosecond lasers. *Appl. Phys. A*. **95**, 643–653 (2009).

35. W. DeSorbo, W. W. Tyler, The specific heat of graphite from 13° to 300°K. *J. Chem. Phys.* **21**, 1660–1663 (1953).

36. E. Pop, V. Varshney, A. K. Roy, Thermal properties of graphene: Fundamentals and applications. *MRS Bull.* **37**, 1273–1281 (2012).

37. *S.R.P. Silva, Properties of Amorphous Carbon, INSPEC, The Institution of Electrical Engineers, London, 2003.*

38. C. Moelle, M. Werner, F. Szücs, D. Wittorf, M. Sellschopp, J. von Borany, H.-J. Fecht, C. Johnston, Specific heat of single-, poly- and nanocrystalline diamond. *Diam. Relat. Mater.* **7**, 499–503 (1998).

39. K. Murata, H. Tanaka, Liquid–liquid transition without macroscopic phase separation in a water– glycerol mixture. *Nat. Mater*. **11**, 436–443 (2012).

40. C. A. Angell, W. J. Sichina, M. Oguni, Heat capacity of water at extremes of supercooling and superheating. *J. Phys. Chem.* **86**, 998–1002 (1982).

41. H. Pathak, A. Späh, N. Esmaeildoost, J. A. Sellberg, K. H. Kim, F. Perakis, K. Amann-Winkel, M. Ladd-Parada, J. Koliyadu, T. J. Lane, C. Yang, H. T. Lemke, A. R. Oggenfuss, P. J. M. Johnson, Y. Deng, S. Zerdane, R. Mankowsky, P. Beaud, A. Nilsson, Enhancement and maximum in the isobaric specific-heat capacity measurements of deeply supercooled water using ultrafast calorimetry. *Proc. Natl. Acad. Sci.* **118**, e2018379118 (2021).

42. Eric W. Lemmon, Ian H. Bell, Marcia L. Huber, and Mark O. McLinden, "Thermophysical Properties of Fluid Systems" in *NIST Chemistry WeBook, NIST Standard Reference Database Number 69* (National Institute of Standards and Technology, Gaithersburg, MD).

43. L. M. Shulman, The heat capacity of water ice in interstellar or interplanetaryconditions. *Astron. Astrophys.* **416**, 187–190 (2004).

44. O. Andersson, H. Suga, Thermal conductivity of low-density amorphous ice. *Solid State Commun*. **91**, 985–988 (1994).

45. S. J. Byrnes, Multilayer optical calculations. *ArXiv160302720 Phys.* (2020) (available at http://arxiv.org/abs/1603.02720).

46. V. Kofman, J. He, I. Loes ten Kate, H. Linnartz, The Refractive Index of Amorphous and Crystalline Water Ice in the UV–vis. *Astrophys. J.* **875**, 131 (2019).

47. X. Wang, Y. P. Chen, D. D. Nolte, Strong anomalous optical dispersion of graphene: complex refractive index measured by Picometrology. *Opt. Express*. **16**, 22105 (2008).

48. G. Rosenblatt, B. Simkhovich, G. Bartal, M. Orenstein, Nonmodal Plasmonics: Controlling the Forced Optical Response of Nanostructures. *Phys. Rev. X*. **10**, 011071 (2020).

49. W. W. Duley, Refractive indices for amorphous carbon. *Astrophys. J.* **287**, 694 (1984).

50. M. Fuchs, E. Dayan, E. Presnov, Evaporative cooling of a ventilated greenhouse rose crop. *Agric. For. Meteorol.* **138**, 203–215 (2006).

51. J. D. Smith, C. D. Cappa, W. S. Drisdell, R. C. Cohen, R. J. Saykally, Raman thermometry measurements of free evaporation from liquid water droplets. *J Am Chem Soc*. **128**, 12892–12898 (2006).

52. B. E. Warren, *X-ray Diffraction* (Dover Publications, 1990).

53. D. T. Bowron, J. L. Finney, A. Hallbrucker, I. Kohl, T. Loerting, E. Mayer, A. K. Soper, The local and intermediate range structures of the five amorphous ices at 80K and ambient pressure: A Faber-Ziman and Bhatia-Thornton analysis. *J. Chem. Phys.* **125**, 194502 (2006).

54. T. Grant, N. Grigorieff, Automatic estimation and correction of anisotropic magnification distortion in electron microscopes. *J. Struct. Biol.* **192**, 204–208 (2015).

55. R. J. List, *Smithsonian Meteorological Tables* (Smithsonian Institution Press, Washington, 1968).